# Metric effects in the space-time with extra-dimensions

Sergey V.Yakovlev

sergey.y@mail.ru

Moscow, April 3, 2012

The aim of this paper is the investigation of scalar and gauge fields on the background metric of Kasner with extra dimensions. Influence of metric effects gets to renormalizing of main parameters of the fields in Lagrangian, and new mass and charge terms are appearing to correspond observed in reality values. Effective masses and charge of the fields are dependant on the additional terms with cosmological constant. For de Sitter stage are obtained expressions for the effective mass and charge of the fields in curved anisotropic space-time.

One of interesting problems in general theory of relativity is investigating of metric effects in the presence of various physical fields. We'll choose as background an anisotropic metric of Edvard Kasner (1922) in multidimensional space-time type with ordinary conditions on metric coefficients [16], [17]. Influence of anisotropy energy of space-time curvature is supposed more high that the density of momentum-energy of the matter in such model. Taking into account only cosmological term, which we'll choose positive, and using exact decisions of Einstein equations for the model of the Universe, we'll try to get some simple metric effects with scalar and electromagnetic field. This problem represents general interest in investigation of physics effects on background of an anisotropic multidimensional Kasner metric with different fields.

For the analysis of the model let's take the background metric in the space-time in the form [16]

$$ds^2 = -dt^2 + e^{2\alpha(t)} e^{2\beta_i(t)} dx^{i\,2},  \qquad (1)$$

where $i$ is running from 1 to $n$, $n$ - the number of space dimensions; scalar field $\phi(x)$ with mass $m$ is describing with the density of Lagrangian

$$L = -\frac{1}{8\pi}(\phi_{,\alpha}\phi^{,\alpha} + (m^2 + \xi R(x))\phi^2),$$

where $R(x)$ is scalar curvature of Ricci of the metric, and $\xi$ is numeral factor. So connection between scalar and gravitational field is represented by the term $\xi R(x)\phi^2$. From here we could get the Klein-Gordon equation for the scalar field

$$\frac{1}{(-g)^{1/2}} \left(\phi^{,\alpha}(-g)^{1/2}\right)_{,\alpha} - (m^2 + \xi R(x))\phi = 0. \qquad (2)$$

The most interested usually are two values of parameter $\xi$: $\xi = 0$, that corresponds to so called minimal connection, and the second one

$$\xi(n) = \frac{1}{4}\frac{n-1}{n},$$

for the case of conformal connection. If $m = 0$ as it's known, equation for the field is invariant under conformal transformations of the metric and scalar field [6], and we're getting the equation of Penrose-Chernikov-Tagirov [6], [14] in the space-time with $n$ space dimensions

$$\frac{1}{(-g)^{1/2}}\left(\phi^{,\alpha}(-g)^{1/2}\right)_{,\alpha} - \frac{1}{4}\frac{n-1}{n}R(x)\phi = 0.$$

In case of mass scalar field we have

$$\ddot{\phi} + n\dot{\alpha}\dot{\phi} - \phi_{,ii}e^{-2\alpha-2\beta_i} + (m^2 + \xi R)\phi = 0. \tag{3}$$

Supposing the uniformity of space, we'll take

$$\phi = \frac{1}{(2\pi)^{n/2}} e^{-n\alpha/2} e^{-i\vec{k}\vec{x}} \chi(t),$$

and get

$$\ddot{\chi} + (m^2 + \xi R + \sum_i k_i^2 e^{-2\alpha(t)-2\beta_i(t)} - \frac{n}{2}\ddot{\alpha} - \frac{n^2}{4}\dot{\alpha}^2)\chi = 0. \tag{4}$$

Einstein equation for the metric with only cosmological term is

$$R_{\mu\nu} - \frac{1}{2}g_{\mu\nu}R = -\Lambda g_{\mu\nu}. \tag{5}$$

From Eq.(5) we have

$$R = 2\frac{n+1}{n-1}\Lambda.$$

For the Einstein's Eq.(5) for the metric (1) we have [16], [17]

$$\dot{\alpha}^2 = \frac{1}{n(n-1)}\left(2\Lambda + \sum_i \dot{\beta}_i^2\right)$$

$$\ddot{\beta}_j + n\dot{\alpha}\dot{\beta}_j = 0 \tag{6}$$

with additional Kasner's condition

$$\sum_{i=1}^n \beta_i = 0.$$

Exact expressions for functions $\alpha, \beta_k$ of metric (1) with $\Lambda > 0$ have the following form [16]

$$\alpha(t) = \frac{1}{n}\ln\left(\frac{B}{\sqrt{2\Lambda}}sh\lambda t\right), \quad \beta_k = \frac{B_k}{B}\sqrt{\frac{n-1}{n}}\ln\left(\frac{B}{\sqrt{2\Lambda}}th(\lambda t/2)\right), \quad \lambda = \sqrt{\frac{2\Lambda n}{n-1}}; \tag{7}$$

here

$$\dot{\beta}_k = B_k e^{-n\alpha}, \quad B^2 \equiv \sum_{k=1}^n B_k^2, \quad B_k = const,$$

and we obtain the final equation for the field $\chi(t)$:

$$\ddot{\chi}+(m^2 -\frac{\lambda^2}{4}+\frac{n+1}{n}\xi\lambda^2 +\sum_i k_i^2 e^{-2\alpha(t)-2\beta_i(t)} +\frac{nB^2}{4(n-1)}e^{-2n\alpha(t)})\chi=0.$$

Imposing on the parameters for simplification additional condition [16], connecting gravitational constant $\Lambda$ with the started dynamic parameter $B$

,

we have finally

$$\ddot{\chi}+(m^2 -\frac{\lambda^2}{4}(1-\frac{1}{sh^2\lambda t}-4\frac{n+1}{n}\xi)+k_i k^i)\chi=0. \tag{8}$$

We see that when $\lambda t \to \infty$ here's appearing the effective mass

$$m_{eff}^2 = m^2 -\frac{\lambda^2}{4}(1-4\frac{n+1}{n}\xi).$$

Taking in the first approximation

$$\chi(t)\approx e^{i\omega_k t},$$

where $\omega_k = k^0$, for the de Sitter stage we have got

$$-k_\mu k^\mu = m_{eff}^2 , \tag{9}$$

That is the effect of changing of mass of scalar particles on the background of Kasner metric. For initial mass in the Lagrangian, in order to get right value and sign of the real mass we could introduce the renormalized value

$$m_{new}^2 = m^2 +\frac{\lambda^2}{4}(1+4\frac{n+1}{n}\xi), \tag{10}$$

so $m^2$ represents the observed value of mass in the de Sitter stage.

Effects like this one, conditioned by metric, are general and getting to renormalizing of parameters of physical fields in curved space-time. In the considered above matter we've investigated effect of renormalizing of mass of the field. The same could be observed also for electromagnetic and other types of fields. It could get to appearing of effective characteristics of the fields on the expanding background metric of space-time [12]. As for the electromagnetic field, we'll see that it could be possible to introduce an effective mass in the Lagrangian and further investigate massive vector field, which mass is cancelled when we're getting expressions for the field functions. The reason of it is the presence of additional metric terms in the field equations.

Consider as an example an electromagnetic field on the backgound (1), (7). Lagrangian of the electromagnetic field is

$$L = -\frac{1}{16\pi}F_{\mu\nu}F^{\mu\nu}, \tag{11}$$

where

$$F_{\mu\nu} = A_{\nu,\mu} - A_{\mu,\nu}, \tag{12}$$

and $A_\mu$ - vector-potential of the electromagnetic field. Maxwell equations in the absence of the currents are written as following

$$F^{\mu\nu}{}_{;\nu} = 0,$$
$$F_{\alpha\beta,\gamma} + F_{\beta\gamma,\alpha} + F_{\gamma\alpha,\beta} = 0. \tag{13}$$

In order to get wave equation for vector-potential, let's substitute Eq.(12) into the first of Eq. (13), and we obtain [6]

$$- A^{\alpha;\beta}{}_\beta + A^{\beta;\alpha}{}_\beta = 0.$$

Recalling that for any vector $B^\mu$

$$B^{\mu;\alpha}{}_\mu = B^\mu{}_{;\mu}{}^\alpha + R^\alpha_\mu B^\mu,$$

where $R^\alpha_\mu$ is Ricci tensor, we have

$$- A^{\alpha;\mu}{}_\mu + A^\mu{}_{;\mu}{}^{;\alpha} + R^\alpha_\mu A^\mu = 0.$$

Finally, using the standard approach of special theory of relativity, imposing an ordinary Lorentz gauge condition

$$A^\mu{}_{;\mu} = 0, \tag{14}$$

we get well known equation [6] with de Rama operator [6]

$$(\Delta_{dR} A)^\mu \equiv - A^{\alpha;\mu}{}_\mu + R^\alpha_\mu A^\mu = 0. \tag{15}$$

Christoffel symbols for the metric (1) have the form

$$\Gamma^i_{i0} = \dot\alpha + \dot\beta_i, \quad \Gamma^0_{ii} = (\dot\alpha + \dot\beta_i) e^{2(\alpha+\beta_i)}, \tag{16}$$

so from Eq.(12) we obtain

$$A^\mu{}_{,\mu} + n\dot\alpha A^0 = 0. \tag{17}$$

For further simplification of investigating of electromagnetic waves in curved space-time, we'll remain only physical degrees of freedom of electromagnetic field and put additional condition

$$A^0 = 0,$$

or from Eq.(17),

$$A^i{}_{,i} = 0. \tag{18}$$

Now let's investigate Eq.(13) for vector potential on the background metric (1), (7). Surely Eq. (13) in flat space-time, when $R^\alpha_\beta = 0$ is getting to the ordinary wave equation of special theory of relativity. Recalling that

$$A^{k;\beta}{}_{;\beta} = A^{k;\beta}{}_{,\beta} + \Gamma^k_{\mu\beta} A^{\mu;\beta} + \Gamma^\beta_{\mu\beta} A^{k;\mu} = \Gamma^k_{\mu\beta} A^{\mu;\beta} + \frac{1}{\sqrt{-g}} (\sqrt{-g} A^{k;\beta})_{,\beta},$$

and using Eq.(14), we have for the space components $A^k$

$$A^{k;\beta}{}_{;\beta} = -\ddot{A}^k - \dot{A}^k((n+2)\dot{\alpha} + 2\dot{\beta}_k) + g^{ii} A^k{}_{,ii} - A^k(\ddot{\alpha} + \ddot{\beta}_k + n\dot{\alpha}(\dot{\alpha} + \dot{\beta}_k)).$$

From Eq.(5) we have

$$R_{\alpha\beta} = \frac{2\Lambda}{n-1} g_{\alpha\beta} = \frac{\lambda^2}{n} g_{\alpha\beta},$$

so Eq.(15) reduces to

$$A^{\alpha;\mu}{}_{;\mu} - \frac{\lambda^2}{n} A^\alpha = 0,$$

and from here we get

$$\ddot{A}^k + \dot{A}^k((n+2)\dot{\alpha} + 2\dot{\beta}_k) - g^{ii} A^k{}_{,ii} + A^k(\ddot{\alpha} + \ddot{\beta}_k + n\dot{\alpha}^2 + n\dot{\alpha}\dot{\beta}_k + \frac{\lambda^2}{n}) = 0.$$

Substituting

$$A^k(x) = \frac{1}{(2\pi)^{n/2}} e^{-(\frac{n+2}{2}\alpha + \beta_n)} a^k(t) e^{i\vec{k}\vec{x}},$$

we get

$$\ddot{a}^k + \omega_k^2(t) a^k = 0, \qquad (19)$$

where

$$\omega_k^2(t) = \vec{k}^2 - \frac{n^2+4}{4}\dot{\alpha}^2 - \frac{n}{2}\ddot{\alpha} - 2\dot{\alpha}\dot{\beta}_k - \dot{\beta}_k^2 + \frac{\lambda^2}{n}, \qquad (20)$$

$$\vec{k}^2 = k_i k^i.$$

From here in the de Sitter stage, when $\lambda t \gg 1$, where $\lambda = \sqrt{\frac{2\Lambda n}{n-1}}$, we have

$$\omega_k^2(t) = \vec{k}^2 - \frac{(n-2)^2}{4n^2} \lambda^2, \qquad (21)$$

or

$$k_\mu k^\mu = \frac{(n-2)^2}{4n^2}\lambda^2 \equiv f^2. \tag{22}$$

So we see that the photon gets an imaginary mass in such an approach. Exact expression for vector-potential could be represented as

$$A^k(x) = \frac{1}{(2\pi)^{n/2}} e^{-(\frac{n+2}{2}\alpha+\beta_\lambda)}(a^k e^{ikx} + a^{k+} e^{-ikx}). \tag{23}$$

For the energy-momentum tensor we have

$$T_{\alpha\beta} = \frac{1}{4\pi} g^{\mu\nu} F_{\alpha\mu} F_{\beta\nu} - \frac{1}{16\pi} F_{\mu\nu} F^{\mu\nu} g_{\alpha\beta}. \tag{24}$$

For zero's component $T_{00}$ from Eq.(24) we obtain the expression

$$T_{00} = \frac{1}{8\pi}(g^{ii} \dot{A}_i^2 + g^{ii} g^{kk} A_{k,i}^2 - A_{,i}^k A_{,k}^i). \tag{25}$$

Substituting Eq. (23) in Eq. (25), for the vacuum state we get

$$<0|T_{00}|0> = \frac{(n-1)}{16\pi^2\pi} e^{-\alpha}(\sum_k \sqrt{k^2-f^2} - e^{-2\alpha}\sum_{i,k} \frac{e^{-(\beta_i+\beta_k)}k_i k_k}{\sqrt{k^2-f^2}} + e^{-2\alpha}\sum_{i,k} \frac{e^{-2\beta_i}k_i^2}{\sqrt{k^2-f^2}}) \tag{26}$$

After regularization over momentum $k_i$ we see that $<0|T_{00}|0> \to 0$ when $t \to \infty$. So for our approximation we obtain

$$<0|T_{00}|0> = 0.$$

The same result's obtaining for the scalar field [17]. We've got zero's density of vacuum energy in the late stage of the expanding of the Universe. Energy-momentum tensor of scalar field for quadratic Lagrangian has the form

$$T_{\mu\nu} = \frac{1}{4\pi}\left(\phi_{,\mu}\phi_{,\nu} - \frac{1}{2}g_{\mu\nu}(\phi_{,\alpha}\phi^{,\alpha} + m^2\phi^2)\right), \tag{27}$$

where instead of $m^2$ we would put for instance $m_{new}^2$ from Eq.(10) for right corresponding with observed value in the de Sitter stage.

Decomposing field along the modes and substituting it into (27), for the vacuum expectation we have

$$T_{00}^{vac} = \langle 0|T_{00}|0\rangle = \frac{1}{8\pi}\sum_k \left\{\left|\dot{u}_k\right|^2 + m^2\left|u_k\right|^2 + e^{-2\alpha(t)-2\beta_i(t)}\left|u_{k,i}\right|^2\right\}. \tag{28}$$

Taking basic modes

$$u_k(t) = \frac{1}{(2\pi)^{n/2}} e^{-n\alpha/2} e^{-i\vec{k}\vec{x}} \chi_k(t),$$

and substituting it into (28), we get

$$T_{00}^{vac} = \frac{e^{-n\alpha(t)}}{(2\pi)^n 8\pi} \sum_k \left\{ \left| \dot{\chi}_k - \frac{n}{2} \dot{\alpha} \chi_k \right|^2 + (m^2 + k_i^2 e^{-2\alpha(t) - 2\beta_i(t)}) |\chi_k|^2 \right\}. \tag{29}$$

For long time approximation in the de Sitter stage, taking state with

$$\chi_k(t) = \frac{1}{\sqrt{2\omega_k}} e^{-i\omega_k t}, \quad \omega_k^2 = m_{eff}^2 + \vec{k}^2,$$

using $2\Lambda = B^2$ and substituting the first of Eq.(6), we get

$$T_{00}^{vac} = \frac{e^{-n\alpha(t)}}{(2\pi)^n 8\pi} \sum_{k<K} \frac{1}{\sqrt{m^2 + \vec{k}^2 - \frac{\lambda^2}{4}}} \left\{ m^2 + \vec{k}^2 + \frac{n(n-1)}{8} \lambda^2 e^{-2n\alpha} \right\} \tag{30}$$

where we've made regularization by offcut for momentum. All anisotropy of $T_{00}^{vac}$ is represented by the second and third terms in brackets.. Because of the divergence we don't have any sense in getting more detailed information concerning $T_{00}^{vac}$. In the de Sitter stage $T_{00}^{vac} \to 0$ when $t \to \infty$, and we could put in our approximation

$$T_{00}^{vac} = 0.$$

As we've seen metric effects lead to renormalization of the main parameters of the models [12]. We take it into account by adding new mass terms into Lagrangian of the fields in order to compensate contributions of metric with cosmological constant in the next way:

$$L_{scalar} = -\frac{1}{8\pi} (\phi_{,\alpha} \phi^{,\alpha} + (m^2 + (\frac{1}{4} + \xi \frac{n+1}{n})\lambda^2)\phi^2), \tag{31}$$

$$L_{em} = -\frac{1}{16\pi} F_{\mu\nu} F^{\mu\nu} - \frac{1}{8\pi} f^2 A_\mu A^\mu, \tag{32}$$

but Lagrangian (32) has lost gauge invariance $A'_\mu = A_\mu + \alpha_{,\mu}$. In Eq.(31) we added for generalization conformal mass term. For the scalar field (31) we're getting decision with real mass $m^2$ and for the vector field (32) massless decision for $A^\mu$ with ordinary $k_\mu k^\mu = 0$ and additional condition $A^i_{,i} = 0$. We should note that all results were obtained for the de Sitter stage, when we could neglect the anisotropy.

As for the complex scalar field interfering with gauge field on the background metric (1), renormalized U(1) gauge invariant Lagrangian of the model has the next form:

$$L = -\frac{1}{16\pi} F_{\mu\nu} F^{\mu\nu} - \frac{1}{8\pi} (\phi_{,\alpha} - ie_{eff} A_\alpha \phi)(\phi^{*,\alpha} + ie_{eff} A^\alpha \phi^*) - \frac{1}{8\pi} \left[ -\mu_{eff}^2 \phi^* \phi + \eta(\phi^* \phi)^2 \right] \tag{33}$$

where we've introduced the effective charge and the mass

$$e_{eff}^2 = e^2 + \frac{2f^2}{\phi_0^2}, \quad \mu_{eff}^2 = \mu^2 + \frac{\lambda^2}{8}. \tag{34}$$

in order to get from Eq.(33) well known form of quadratic Lagrangian (35). Renormalizing of mass leads to renormalizing of charge of gauge field on the background of Kasner type. Main state of the scalar field is

$$\phi_{vac} = \frac{1}{\sqrt{2}} \phi_0 = \frac{\sqrt{\mu^2 + \frac{\lambda^2}{8}}}{\sqrt{2\eta}},$$

and $\phi$ could be represented as

$$\phi = \frac{1}{\sqrt{2}}(\phi_0 + \chi(x) + i\theta(x)).$$

Quadratic Lagrangian we're obtaining in an ordinary way,

$$L^{(2)} = -\frac{1}{16\pi} B_{\mu\nu} B^{\mu\nu} - \frac{1}{16\pi} e_{eff}^2 \phi_0^2 B_\nu B^\nu - \frac{1}{16\pi}(\chi_{,\nu}\chi^\nu + 2\mu_{eff}^2 \chi^2) + \frac{\mu^4}{32\pi\eta},$$

(35)

where as usually

$$B_\mu = A_\mu - \frac{1}{e_{eff}\phi_0}\theta_{,\mu}.$$

Using in Eq.(33) effective charge $e_{eff}$ of gauge field we're getting an ordinary Eq.(35) for the Lagrangian (33), but we're taking into account additional metric terms in the masses of all fields according to Eq.(31), (32). Mass of vector field $B^\mu$ is becoming equal

$$m_V^2 = \frac{1}{2} e_{eff}^2 \phi_0^2 = f^2 + \frac{1}{2} e^2 \phi_0^2 = \frac{(n-2)^2}{4n^2}\lambda^2 + \frac{e^2}{2\eta}(\mu^2 + \frac{\lambda^2}{8}). \tag{36}$$

In the de Sitter stage the term with $f^2$ will be cancelled, so for this case we'll have for the pure mass of vector field

$$m_V^2 = \frac{1}{2} e^2 \phi_0^2 = \frac{e^2}{2\eta}(\mu^2 + \frac{\lambda^2}{8}). \tag{37}$$

Mass of scalar field $\chi$ in Lagrangian (35) equals

$$m_\chi^2 = 2\mu_{eff}^2 = 2\mu^2 + \frac{\lambda^2}{4},$$

and it's real mass in the de Sitter stage

$$m_\chi^2 = 2\mu^2.$$

For generalization we would introduce term with $\mathcal{R}(x)|\phi|^2$ into Lagrangian (33), and after that we would get additional terms into all masses of the fields. Introducing of new charge of gauge

field in Eq.(33) is needed for getting the right value of mass (37) of the vector field in the de Sitter stage. Whole procedure of changing of masses and charge in Lagrangians is introduced for clarification of what parameters in Lagrangians we could observe in reality.

Thus we've obtained effective masses and charge of the fields in Kazner metric and renormalized expressions for Lagrangians of scalar and gauge fields, which yield us right values of masses for long times of expanding of the Universe. In general by taking into account metric effects for the fields we could obtain new effective expressions not only for masses and charge but for other field's characteristics, not disappearing in the flat world, that is important in investigating of dynamic of the Universe [12]. Metric effects play fundamental role in estimation of dynamic variables, and there exist some cases when, for example, spin of the field is determining by only interaction of the field with metric field even when gravity is absent, but surely not the metric [12]. That new variables should be included into expressions that have to be used in the right-hand side of Einstein equation [18], but that needs an additional investigation.